\documentclass{aastex62}

\newcommand{{\HII}}{H\,{\sc ii}}

\newcommand{\um}{\,$\mu$m}
\newcommand{\kms}{\,km\,s$^{-1}$}

\definecolor{mygreen}{rgb}{.05, .6, .01}
\definecolor{roxa}{rgb}{.5, .02, .6}
\definecolor{aqua}{rgb}{0.0, 1.0, 1.0}
\definecolor{azure}{rgb}{0.0, 0.5, 1.0}
\definecolor{bondiblue}{rgb}{0.0, .58, .71}
\newcommand{\harotwo}{Haro~2}
 
\accepted{MNRAS 2020 March 2}

\shorttitle{SMA Observations of Haro 2}
\shortauthors{Beck et al.}
\watermark{\today}

\begin{document}

\title{SMA Observations of Haro 2: Molecular Gas around a Hot Superbubble }
\correspondingauthor{Sara Beck}
\email{becksarac@gmail.com}

\author[0000-0002-5770-8494]{Sara C. Beck}
\affil{School of Physics and Astronomy, 
Tel Aviv University, Ramat Aviv, Israel}

\author{Pei-Ying Hsieh}
\affil{Academica Sinica, Institute of Astronomy and Astrophysics,
P.O. Box 23-141, Taipei 10617, Taiwan}

\author[0000-0003-4625-2951]{Jean Turner}
\affil{Department of Physics and Astronomy, UCLA, Los Angeles, CA 90095-1547}

\begin{abstract} 
\harotwo\ ,  a nearby dwarf starburst dwarf galaxy with  strong Ly~$\alpha$ emission, hosts a starburst that has created outflows and filaments.  The clear evidence for 
galactic outflow makes it an ideal candidate
for studying  the role of molecular gas in feedback processes in a dwarf galaxy.  
We observed CO(2-1) in \harotwo\  at the Submillimeter Array in the compact and extended configurations, and have mapped the molecular emission with velocity resolution 4.1\kms~ and spatial resolution $2.0\times1.6$\arcsec~. With this significant increase of resolution over previous measurements we see that the molecular gas comprises two components: bright clumps associated with the embedded star clusters of the starburst, and fainter extended emission east of the starburst region. The extended emission coincides with an X-ray bubble and has the kinematic signatures of an outflowing cone or of an expanding  shell or bubble; the velocity range is $\sim35$\kms~.   We suggest that the starburst winds that created the X-Ray bubble have entrained molecular gas, and that the apparent velocity gradient at an angle to the photometric axis is an artifact caused by the  outflow. The molecular and X-ray activity is on the east of the galaxy and the  ionized outflow and optical filaments are west; their relationship is not clear.  
\end{abstract}


\section{Introduction} \label{sec:intro}
Blue compact dwarf (BCD) galaxies form stars in environments quite different 
 from gas-rich spirals or luminous infrared galaxies. 
BCDs are often rich in
 atomic gas but it is difficult to study their molecular components; the CO emission is usually weak and BCDs are often of low metallicity which complicates the interpretation of the emission. 
 The origin of the molecular gas from which the stars
are forming is thus unclear. Perhaps star-forming molecular clouds form from their reservoirs of atomic gas; alternatively, molecular gas 
could be acquired from other galaxies via accretion or merger. Another urgent question is how starburst feedback operates on
molecular gas. 
 The gravitational wells of dwarf galaxies are not as deep as spirals' and it is easier for them to lose gas to winds. However,
  molecular gas has greater cooling capacity than atomic, which can cause galactic winds to fail and inhibit gas dispersal. This can
 have ramifications for the escape of ionizing radiation from the starburst.
 
 \harotwo\ (Mrk 33, Arp 233, UGC~5720) is one of the best local targets in which to study
the processes of star formation fueling and feedback in a dwarf
galaxy. 
One of the most luminous  BCDs, \harotwo\ 
has a very blue nucleus \citep{1956BOTT....2n...8H} is of moderate metallicity ($Z\sim Z_\odot/3$), with strong ultraviolet and optical emission lines and WR features 
\citep{1985A&A...142..411K,1986ApJ...309...59L,1993ApJS...86....5K}, and
bright radio continuum emission \citep{2000AJ....120..244B,2011AJ....141..125A}, indicating very recent ($<10$~Myr) star formation.
\harotwo\ is one of the closest  Lyman $\alpha$ emitting galaxies;
Its strong, blue-shifted Ly $\alpha$ line 
is probably due to a outflow of ionized gas at $\approx200$\kms~ with respect to the galaxy  
 \citep{1995A&A...301...18L,2001AJ....121..740M}.  
 The ionized outflow is presumably driven by the massive stars in the starburst.  
  \citet{2001MNRAS.327..385S} observed the soft X--ray emission of \harotwo\ with HRI on ROSAT with 1.5\arcsec~pixels and saw  ``an extended, complex shell-like morphology"; they found hard emission concentrated in three point sources as well as the widely distributed soft emission. 
 CO(1--0),CO(2--1), and CO(3--2) lines have been detected with single dishes \citep{
 1988A&A...205...41A,1992A&A...265...19S,2000MNRAS.317..649B,2001AJ....121..740M,2010ApJ...724.1336M,1995A&A...295..599I}.  
  \harotwo\ is comparatively 
weak in low J  lines of CO relative to the brightness of the starburst, which is common for BCDs.

 With sensitive array observations, CO  can be mapped at the scales of individual
 giant molecular clouds (GMCs) in nearby galaxies, allowing the star formation process to be studied at the
 cluster scale.  The first interferometric maps of CO and HI in \harotwo\ were by \citet{2004AJ....127..264B}; their map of CO(1--0) 
 revealed unusual halo-like
 emission around the galaxy. They found the molecular gas to have a velocity gradient almost due north-south, at an angle $\sim40^o$ to the photometric major axis of the galaxy, and suggested that the photometric and kinematic axes are misaligned.  
 
This motivated us to observe \harotwo\ in a higher transition; CO lines from high J are  often stronger than CO(1-0) in BCDs, especially in regions of star formation \citep{1995A&A...295..599I,2001AJ....121..740M}.  We accordingly observed
\harotwo\ in the CO(2-1) line with the Submillimeter Array (SMA) on Mauna Kea in both the extended and compact configurations, giving spatial resolution better than 2\arcsec~ and velocity resolution of $\approx4$\kms~ over the star forming region (observational parameters are in Table 1).  We combine this data cube with archival radio continuum and optical images for a full picture of molecular gas kinematics in the center of \harotwo\ which is very different from the earlier  results. 

\section{Observations}

We
observed the CO(2-1) line in both the extended and compact configurations of the Submillimeter Array (SMA).  The compact configuration, which has a maximum baseline of 70 m, was used on 13/6/2015 and the extended configuration with maximum baseline  220 m on 19/4/2016. 
Callisto was the absolute flux calibrator.  Full parameters of the observations and the range of spatial resolutions achieved are in Table 1.  The data was calibrated in MIRIAD and imaged  with AIPS and CASA 4.7.0.  Combining the extended and the compact configuration data created a data cube with spatial beam of $1.96\times1.61$\arcsec~, almost as high as the extended configuration, but as sensitive as the compact configuration to large-scale structures; the largest structure that can be mapped is $\sim210$\arcsec, corresponding to  22 kpc at the galaxy's distance of 21 Mpc.    Final noise levels in the individual combined channel maps are 17 $\rm mJy~bm^{-1}$. The
integrated intensity (moment 0)  image was constructed from all channels stronger than  $>1.5\sigma$. The intensity
weighted mean velocity and dispersion maps (first and second moments)  were constructed from emission $>4\sigma$. 
 \citet{2000MNRAS.317..649B} obtained a peak  $I_{co}=7.6\pm0.3 K$\kms~ for \harotwo\ at the IRAM 30-m telescope, which gives a total flux density of $\approx84$-125 Jy\kms~, depending on the true source size.  The SMA observations find a total flux $S(CO(2-1))$ of 110 Jy\kms~,  consistent with the single-dish measurements as far as can be determined. 
    
  \section{SMA Observations of  CO(2--1): Distribution of Emission within Haro~2}
  
  \begin{figure}[ht]
\begin{center}
\includegraphics*[width=3.5in]{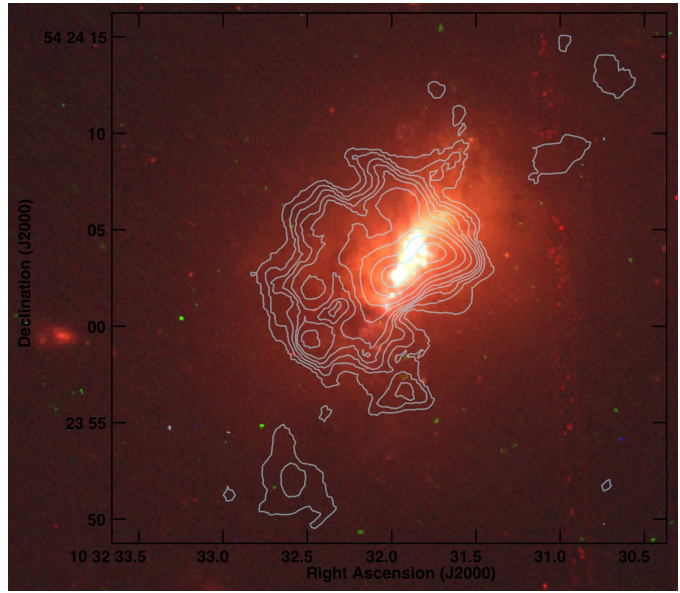}
\caption{Zero moment (total intensity) map of the CO(2-1) emission in \harotwo\ in contours, superimposed on an 814nm image from the Hubble Legacy Archive.  The contour interval is 1800, starting at 1250; units are (Jy/bm)(m/s).  The alignment of the images is estimated to be within 0.5\arcsec.    }
\end{center}
\end{figure}

\subsection{Basic Structure: Starburst Clumps and an Extended Northeast Wing}
The CO(2-1) total integrated intensity map of \harotwo\ is displayed in Figure 1, 
as contours on an 814nm HST image.  The total extent of the CO emission is about 1.5--2 kpc, about the size
of the optical image. The brightest CO coincides with the brightest starburst region and there is extended CO emission east and north of the starburst. The edges of the extended CO lie near weak dust lanes, indicating that the extended CO is on the near side of the galaxy. These features are also seen in the CO(1-0) images of \citet{2004AJ....127..264B}.  We do not see CO(2-1) west of the starburst nucleus where \citet{2000A&A...359..493M} found $H\alpha$ filaments.

The CO(2-1) flux, $S_{21}$, is distributed with $\approx 50$ Jy\kms\ in the central starburst region and 50-60 Jy\kms\ in the extended north-east emission.   \citet{2004AJ....127..264B} find a CO(1-0) flux, $S_{10}$, of $\approx17.6$~Jy\kms\  in the central region and 6-7 Jy\kms\ in the extended. The $S_{21}/S_{10}$ ratio of 2.9  in the central region is typical of compact galaxies \citep{2005A&A...438..855I} and thermally excited gas at $T\approx10 K$.   The $S_{21}/S_{10}$ flux ratio of the extended blue emission is 7-9,  but this ratio simply reflects the fact that the CO(1-0) emission falls below   the  $1\sigma$ intensity cut of the CO(1-0) map at 100 mJy/bm\kms\ .   This corresponds, for a typical $S_{21}/S_{10}$ ratio of 3, to an rms almost 5 times higher than the $\approx65$ mJy/bm\kms\ of our more sensitive maps.  We do not expect to see the extended blue emission detected by the SMA  in the \citet{2004AJ....127..264B} maps.

We can estimate the total mass in the extended emission from the observed 50-60 Jy\kms\  by assuming $S_{21}/S_{10}\approx3$  and $\rm X_{CO}$, the conversion factor giving $M_{H_2}$ from CO flux. For  \citet{2004AJ....127..264B}'s $\rm X_{CO}$ value of 
$2\times10^{20}~\rm cm^{-2}\,(K\, km \,s^{-1})^{-1}$, there is $5\times10^7M_\odot$ in the extended emission region, and $1\times10^8 M_\odot$  for $\rm X_{CO}=4\times10^{20}~\rm cm^{-2}\,(K\, km \,s^{-1})^{-1}$.

\begin{figure}[h]
\includegraphics[scale=0.7]{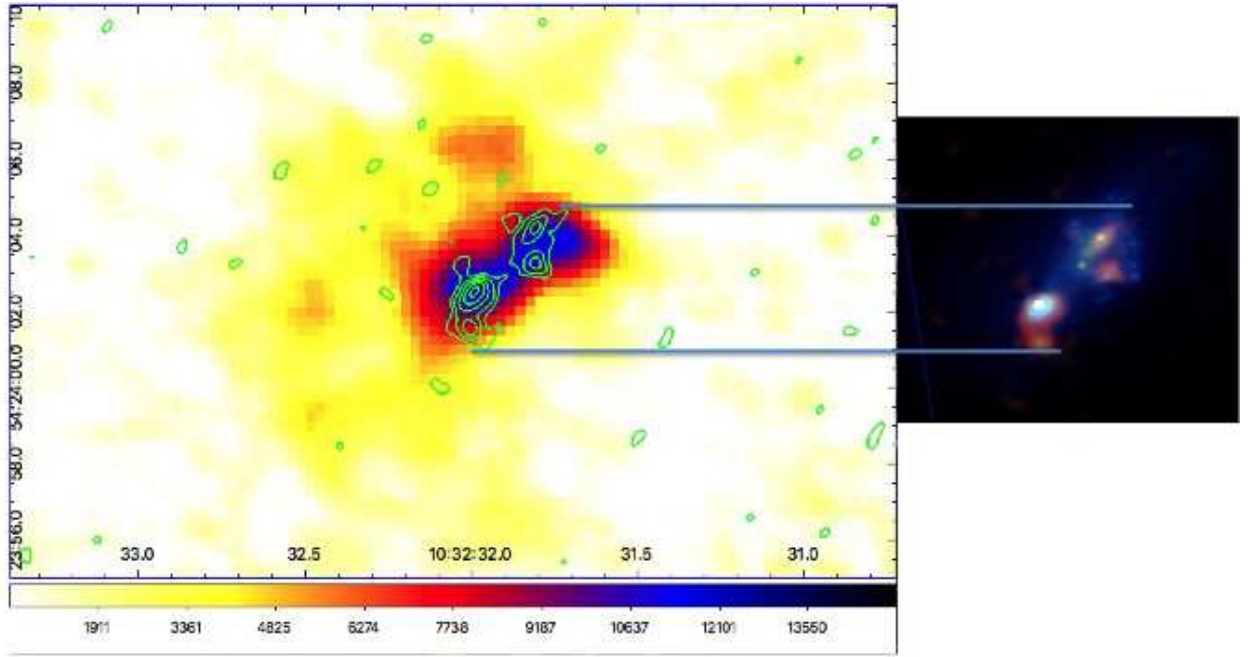}
\caption{The CO(2-1), radio and near-infared emission of \harotwo\~.  The left panel displays the integrated intensity (moment 0) of CO(2-1) in color, with the 6 cm radio continuum overlaid in green contours.   The color wedge units are Jy/bm m/s and contours are multiples of $1.5\times10^{-4}$ Jy/bm from the base of $2.5\times10^{-4}$ Jy/bm.  The left panel is cropped to the region of star-forming clumps as marked by the lines  and shows the  6 cm radio continuum in red and the 1.6\um~ emission (from NICMOS) in blue. } 
\end{figure}

\begin{figure}[h]

\begin{center}
\includegraphics[scale=0.5]{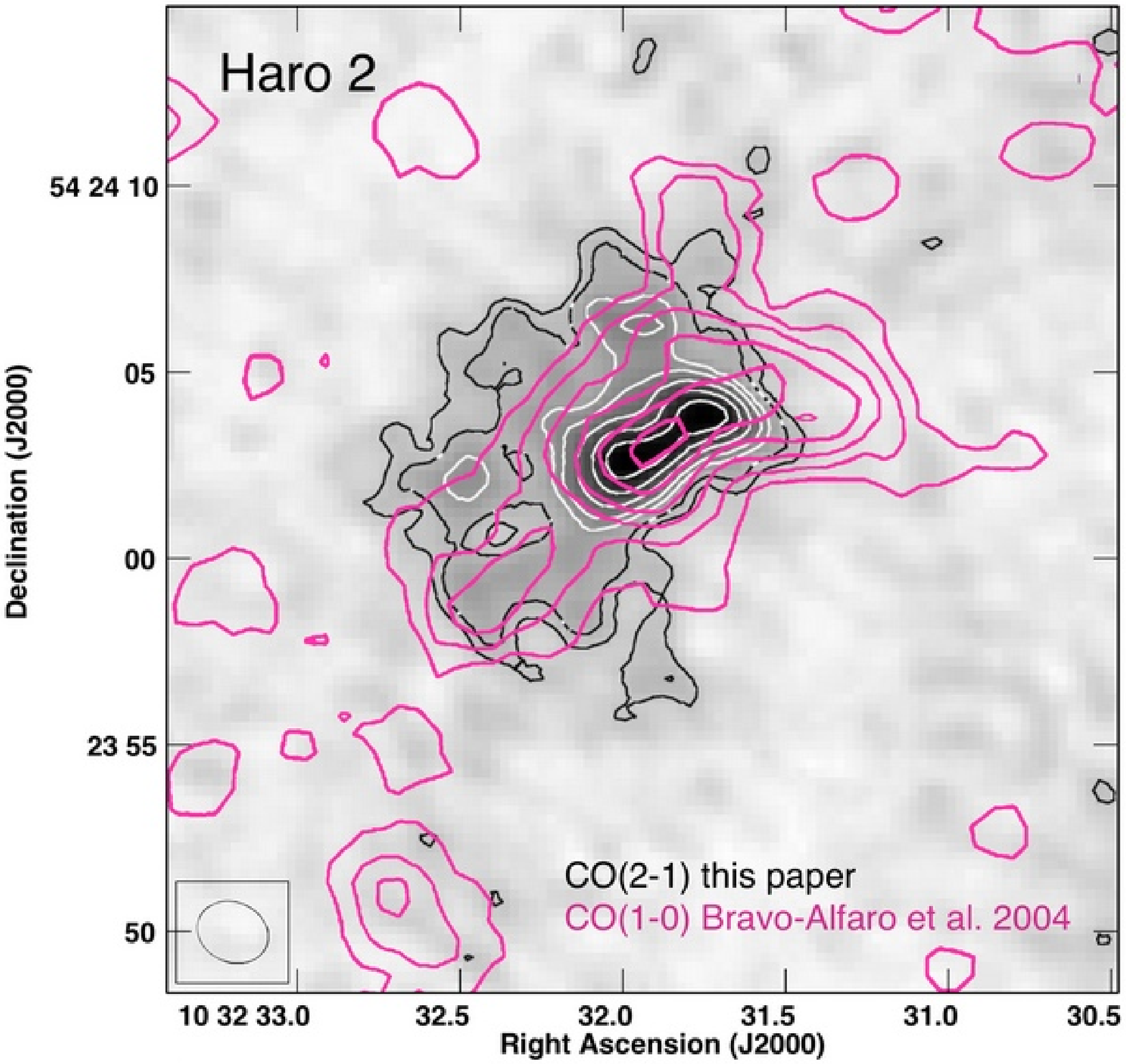}
\caption{The CO(2-1) emission in grayscale and black and white contours together with the (1-0) emission from \citet{2004AJ....127..264B} in colored contours. The (2-1) is enhanced relative to the (1-0) north and east of the starburst region. }
\end{center}
\end{figure}
Figure 3 compares the CO(1-0) and CO(2-1) maps and shows the CO(2-1) extending north and east of the CO(1-0) emission.    This will be discussed further in relation to the X-ray emission below.

\citet{2004AJ....127..264B} report `possible' CO(1-0) concentrations south and northwest of the main emission region in a map made with a  $3.3\times2.6$\arcsec~ beam .  These features are very weak at the full spatial resolution; tapering the beam to $\approx3$\arcsec~  shows emission at the $2-4\sigma$ per beam level northwest and south east of the main galaxy. These possible features appear in the moment maps of Figure 5. We believe them to be real, especially the extended northwest feature which covers several beams, but the signal-to-noise at this time is too low for meaningful study and they will not be discussed further here.  

  The starburst in \harotwo\ has been mapped in the radio continuum by  \citet{2000AJ....120..244B} and \citet{2011AJ....141..125A} with beams of 2\arcsec\ and 1\arcsec\ respectively; at this resolution the emission appears as  two bright regions.  An archival VLA image  from the AY35 program, shown in Figure 2,  has resolution $0.58\times0.47$\arcsec\ and shows that each of the bright regions breaks up into a few clumps.    Figure 2 shows the CO(2-1) moment 0 overlaid with the high resolution radio map and demonstrates that the strongest CO emission mostly coincides with the two bright starburst clumps.  In the right panel of Figure 2 we display the radio continuum and the 1.6\um~ continuum mapped by NICMOS \footnote{As astrometry from the NICMOS headers is known to be imprecise at the 1-2\arcsec~level  \citep{2012A&A...546A..65O},  we shifted the NICMOS image $\sim0.5$\arcsec north-east to bring the strong southern infrared source into coincidence with the bright radio clump.}

\section{SMA Observations: the Kinematics of the Molecular Gas}

The kinematics of gas in \harotwo\ are of particular interest because it is one of the closest Lyman $\alpha$ emitting galaxies.  Since
\harotwo\ does not have a particularly low metallicity, it has
been suggested \citep{1995A&A...301...18L} 
that the escape of Lyman continuum photons is due to gas kinematics, and indeed there is a suggestion 
of outflow in the ionized gas \citep{1997A&A...326..929L}. How this might compare to the kinematics of the molecular gas is an important question.
Furthermore, previous observations have found signs of unusual and non-rotational kinematics in  \harotwo\ .  \citet{2004AJ....127..264B} and \citet{2006RMxAA..42..261B}  report that in both HI and CO  the photometric and kinematic axes appear to be misaligned, with little velocity gradient along the major axis and a strong gradient at an angle of $\sim 40^o$ to it; similar results are reported for H$\alpha$ by \citet{2002A&A...391..487P}.  These results have been taken as signs of a past accretion or merger event (\citet{2004AJ....127..264B} and \citet{2006RMxAA..42..261B})  However, the CO data cube on which these arguments were based only covered velocities of 1378 to 1503\kms~ and had beam size $3.2\times2.6$\arcsec.  The SMA data with its better spatial resolution and more complete velocity coverage presents a somewhat different picture, which we now discuss.  

Figure 4 displays samples of channel maps of CO(2--1) emission (a complete set of channel maps is in the Appendix).  The CO(2-1) emission in \harotwo\ spans a total velocity range of $\approx200$\kms~ FWZI,  from about 1380 to 1580 \kms, with a centroid of 1470\kms~ ( \citet{2004AJ....127..264B} found 1440\kms~ as the center of the HI line).   
Figure 5 displays the first moment (intensity-weighted peak velocity) and second moment (dispersion) of the CO(2-1) line, with the radio continuum contours superimposed to show the location of the young star clusters and the photometric axis of the galaxy. 
These displays  of the kinematics show:  
\begin{itemize}
\item Along the axis defined by the positions of the young star clusters there is a smooth velocity gradient from  $\approx1465$\kms~ on the north to $\approx1510$\kms~ on the south.
\item The extended emission north and east of the starburst is blue-shifted relative to the starburst and has velocities between $\approx1420$\kms~ and $\approx1460$\kms~. 
\item The velocity field of the extended blue emission is complex. 
\item Over most of the region of extended blue emission the dispersion is high. 
\end{itemize} 
We conclude that the galaxy holds two distinct kinematic regions: the starburst lying along the photometric axis,  and the blue emission north-east of the main body of the galaxy. We now discuss each component.
 
\subsection{Molecular Gas Kinematics in the Starburst}
The channel maps of Figure 3 clearly show two velocity components associated with the starburst: one at
$\sim1510$\kms\  on the southeast star-forming clump and another component at  $\sim1470$\kms\ on the north clump, giving
a centroid of 1490\kms.  How do these velocities compare to those seen at optical wavelengths?  The $H\alpha$ spectra of \citet{2000A&A...359..493M}  found velocities of $\approx1430$, $\approx1450$ and $\approx1465$ on their $H\alpha$ knots 1,2,3 respectively.  They do not give absolute coordinates for their positions, but from their offsets and by comparing the H$\alpha$ image to the near-infared, we conclude that their $H\alpha$ knots 1 and 2 are north of the radio emission region and  {\it not} associated with the star forming clumps. Their knot 3 corresponds to the the northern radio source and agrees with it in velocity.  That the southern radio source has no associated $H\alpha$ is a sign of high local extinction, which plays an important role in the appearance of this galaxy's core.    

The central panel of Figure 6 shows the velocity as a function of position (Position-Velocity Diagram of PVD)  along the main body of the galaxy as defined by the radio clumps. The velocity shift between the two sources is consistent with the gradient set by the $H\alpha$ clumps, further north along the same axis.   The patch of  high dispersion with $\sigma\approx$35\kms~ that appears in Figure 5 between the two radio concentrations shows where emission of both groups of sources enter the beam.   

 
   \begin{figure}[h]
\includegraphics*[scale=0.9]{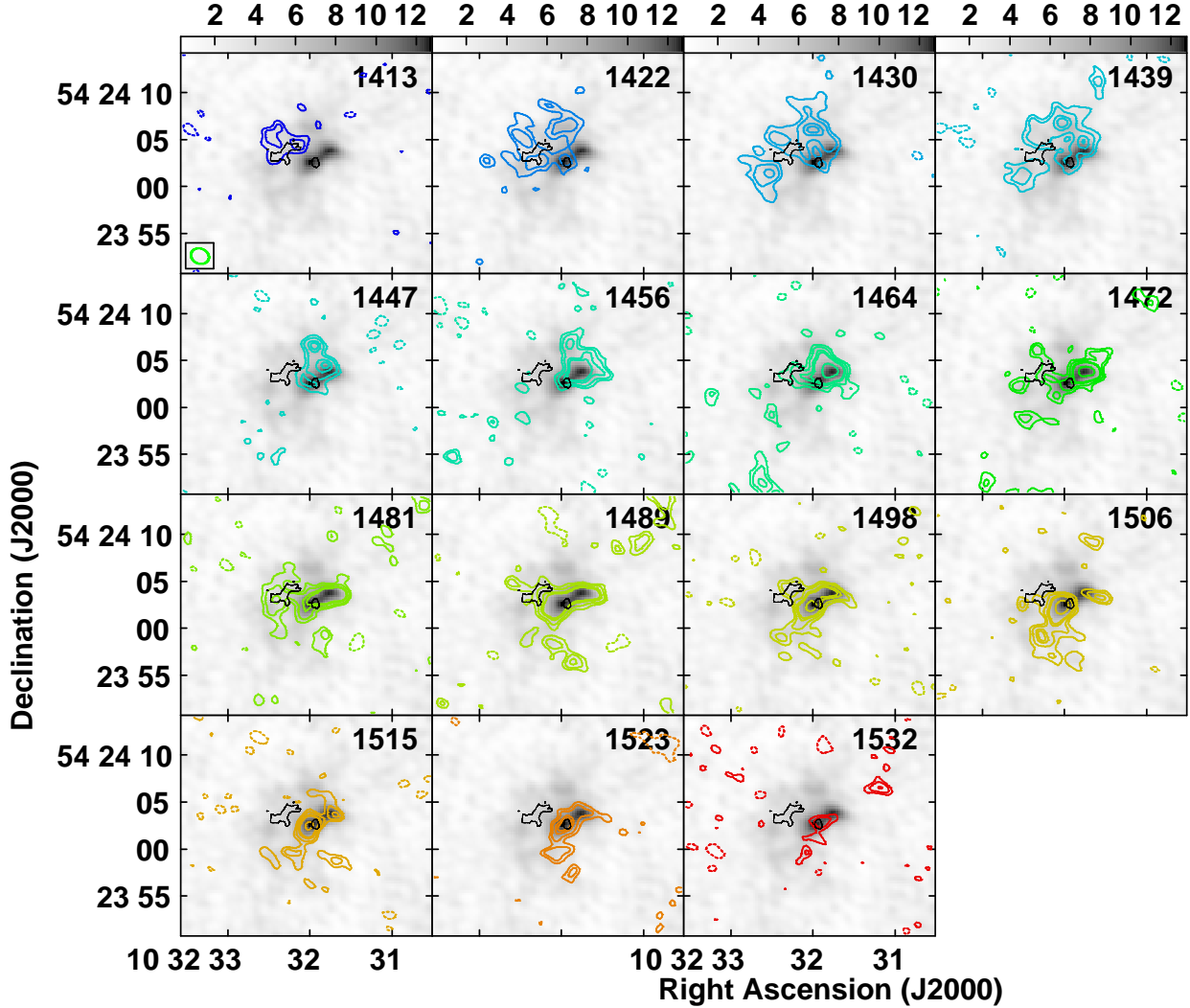}
\caption{Contours of the CO(2-1) line emission in a 4\kms~ channel at the velocity shown in each panel, overlaid on the total intensity map in greyscale. Panels are labelled by velocity and contour levels are  $4\times10^{-2}\times2^{n/2}$ Jy/bm.  Every other  channel in the range of strongest emission are shown here; the full set is in the Appendix. The thin black contour in each panel is the contour of highest 
velocity dispersion shown in Figure 5.   }
\end{figure}

\begin{figure}[h]
\plottwo{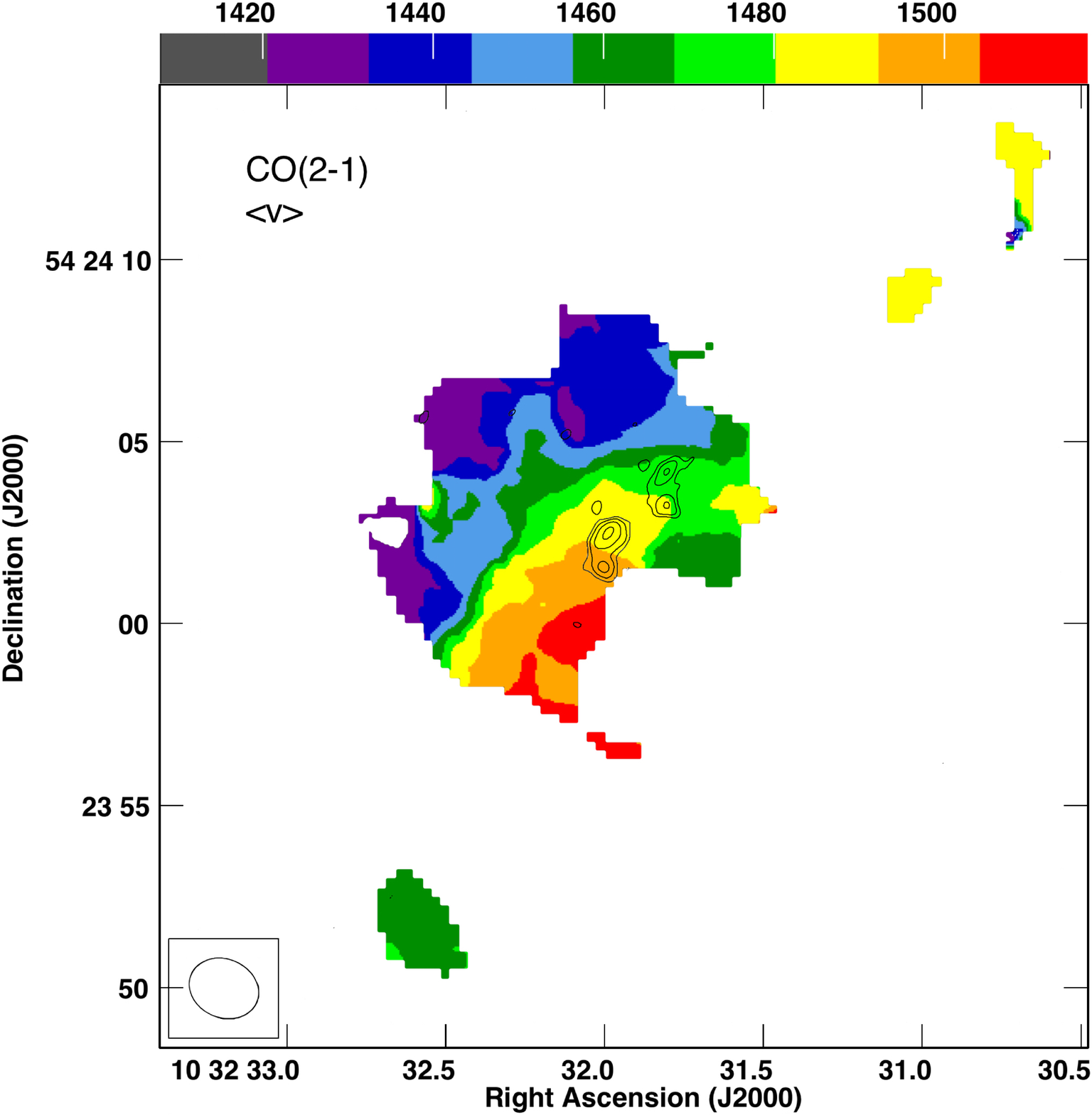}{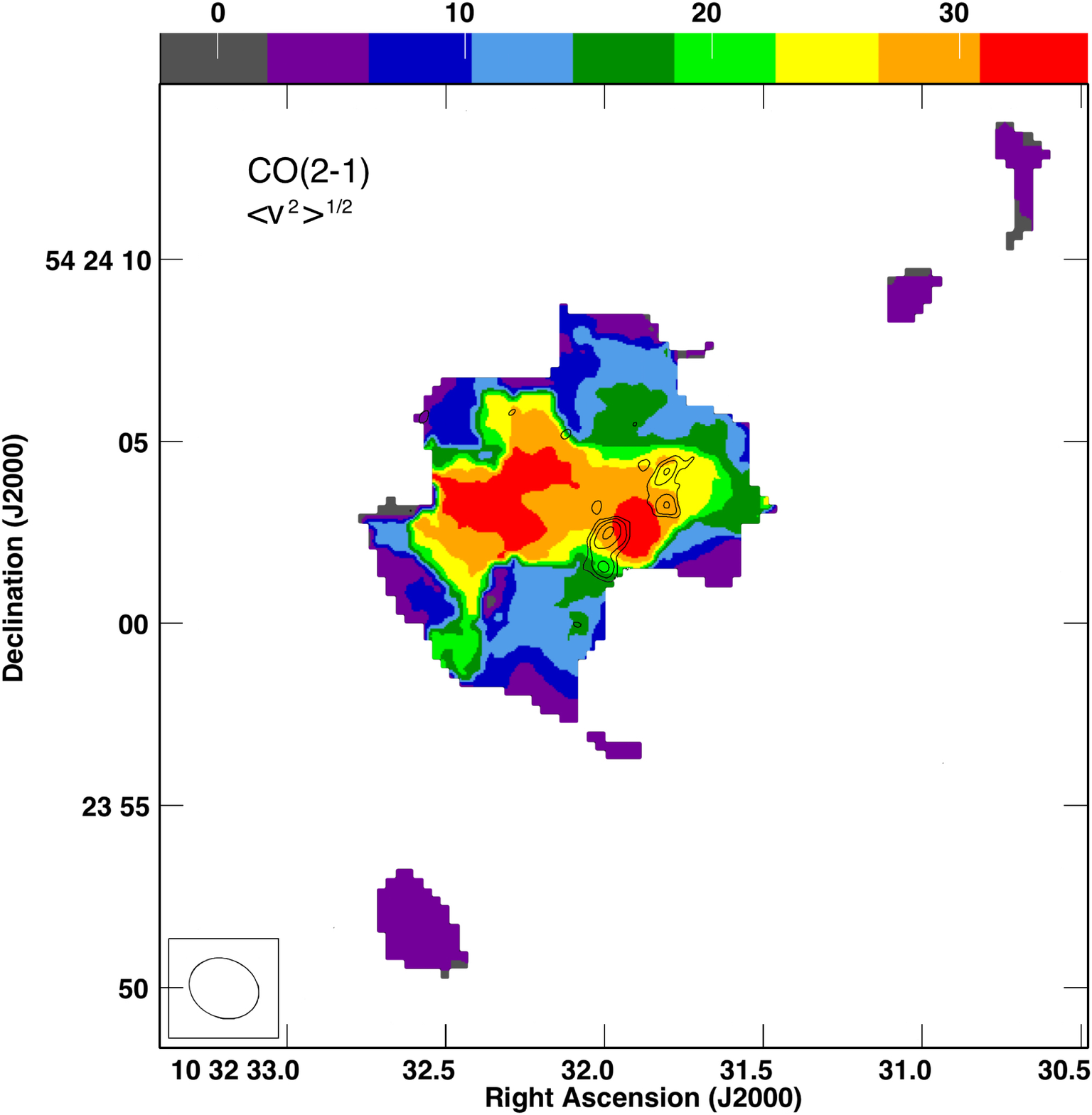} 
\caption{Intensity-weighted mean velocity (left) and velocity dispersion (right) of  CO(2--1) in \harotwo\ .   The units of the color wedges are \kms~.  The radio contours of Figure 2 are superimposed on the moment maps. }
\end{figure}

\subsection{Molecular Gas North East of the Starburst:  A CO Outflow or Expanding Bubble}
 
The emission north-east of the starburst   extends over  $\sim$ 1.5-2 kpc and 1400-1500\kms\  in velocity.  The first and second moment maps of Figure 5 and the channel maps in Figure 4 and the Appendix show that the extended eastern emission is entirely blue of the main starburst; the blue velocities of the extended eastern emission have no counterpart in the optical spectra.  The eastern emission is concentrated at two velocities  $\approx45$\kms\  apart; these appear in the channel maps of Figure 4 as two `horns' extending east from the starburst region.   These features appear clearly in Position-Velocity Diagrams (PVDs) in Figure 6.  In short,  the channel maps, PVD and line profiles of the molecular gas east of the starburst all demonstrate the kinematic signatures of a body of gas flowing away from the galaxy with velocity $\sim35-45$\kms.   

Two types of motions that could produce this kinematic signatures are 1) an expanding shell or bubble, 2) outflow cones based in the starburst.   In a true bubble expanding from a central point,  the highest velocities and the greatest velocity dispersion will be seen in the center. In the \harotwo\ extended emission region the most extreme gas velocities and the region of highest velocity dispersion, shown as a thin black contour on the channel maps, are not in the center of the emission but closer to the starburst.  This may argue that the blue velocities show outflows from the clusters. However, the gas spatial distribution is patchy and if there is a shell it is very incomplete and difficult to see. It may not be possible with the current data to distinguish clearly  between expansion of a  partial shell and an set of outflows.  We refer below to this component of molecular gas as a 'bubble' because it coincides with a well-known X-ray bubble, with the caveat that the molecular gas may not have identical morphology. 
 \begin{figure}[h]$
\begin{array}{ccc}
\includegraphics*[scale=0.25]{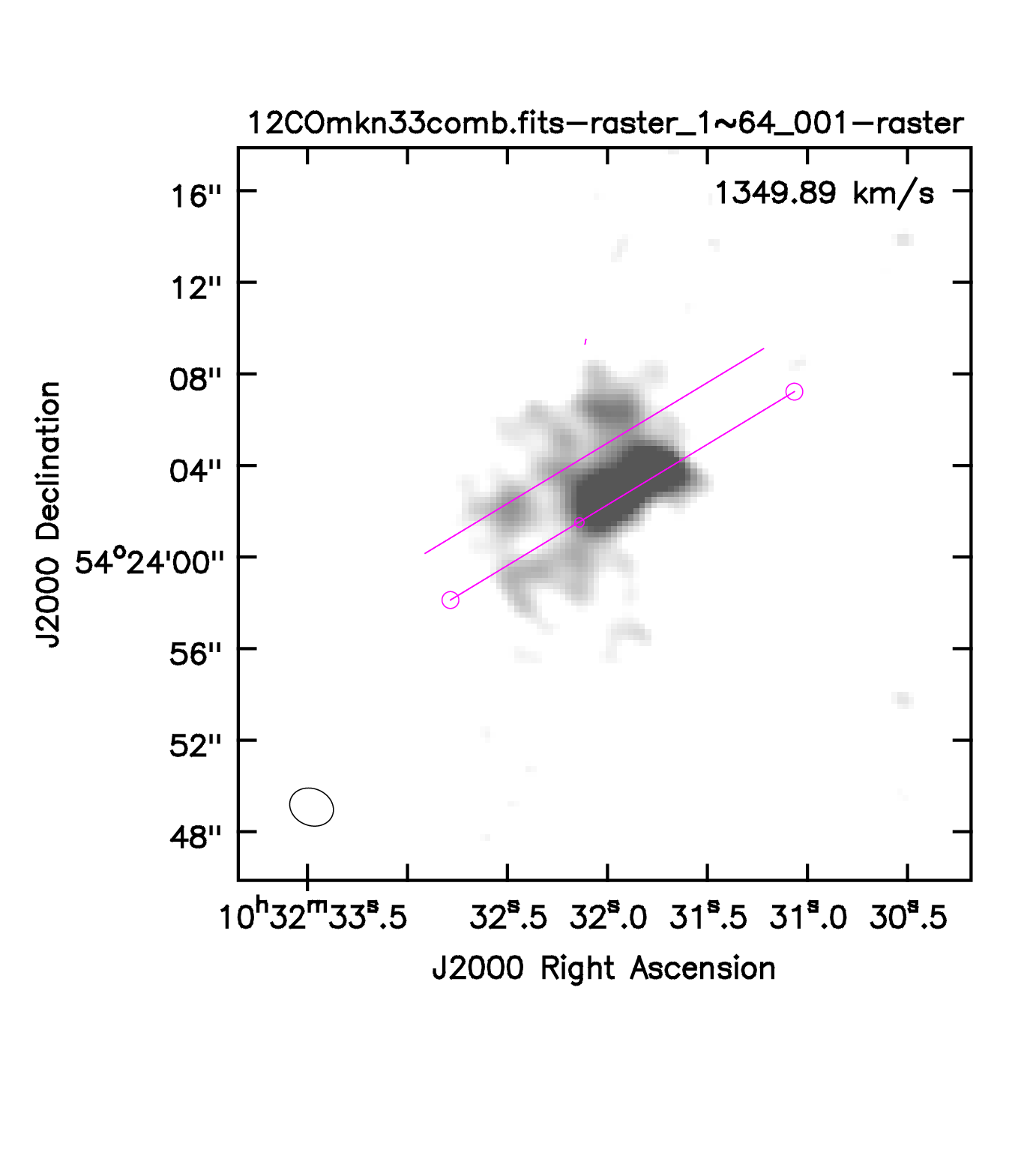} & \includegraphics*[scale=0.25]{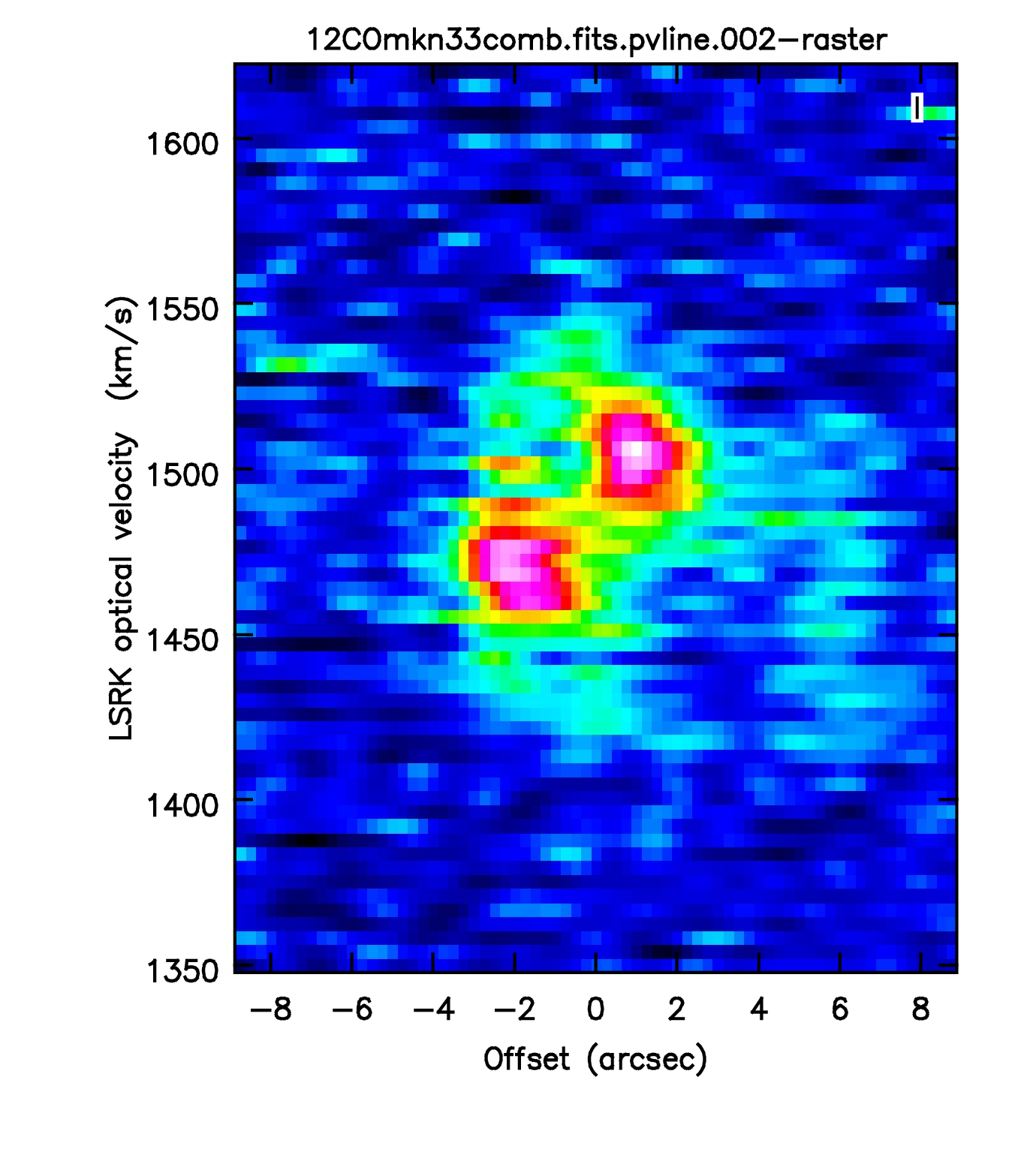} & \includegraphics*[scale=0.25]{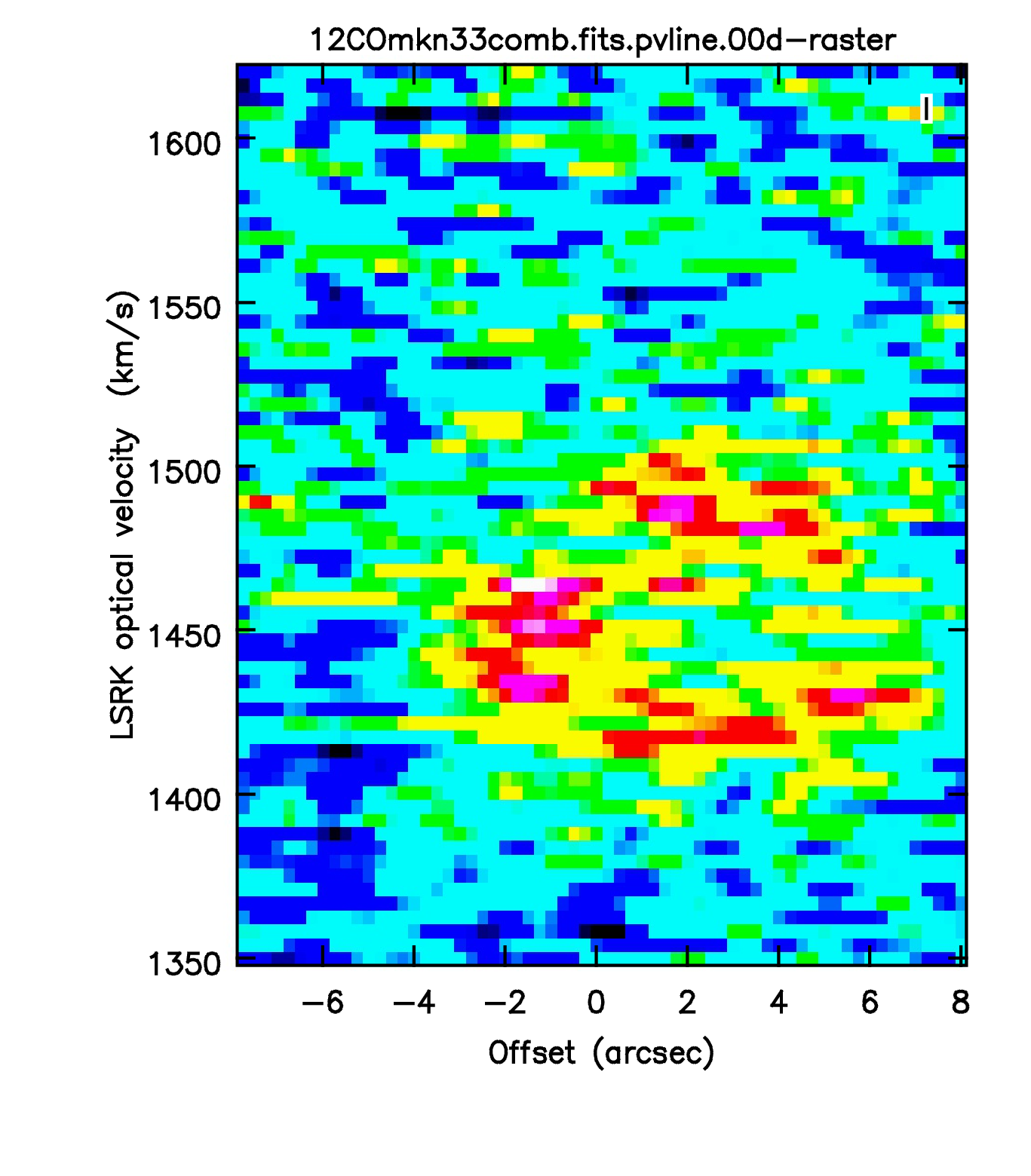}   \\
\end{array}$
\caption{Position-Velocity diagrams of the CO(2-1) emission. The left panel shows the slices on which the velocities were measured, overlaid on the moment 0 total emission map in grayscale. The middle panel is the PVD on the  lower line, through the main clusters, and the right panel is the PVD on the upper line through the bubble.  Negative offsets are east. }
\end{figure}

 \subsubsection{The Molecular Outflow and the X-Ray Emission}
The large-scale activity reflected in the molecular outflow appears as well in the high energy regime. 
\citet{2001MNRAS.327..385S} observed \harotwo\  with HRI on ROSAT with a $\sim5$\arcsec~ beam and 1.5\arcsec~pixels and saw  "an extended, complex shell-like morphology" .   \citet{2012A&A...546A..65O} observed with CHANDRA over a wider energy range and with 0.49\arcsec~ pixels.  Their image of the soft  (0.2-1.5 KeV) emission is very similar to the ROSAT result; they find in addition 3 point-like sources of hard (2.5-8.0 KeV) X-rays.  How does the structure correspond to the molecular gas?    In Figure 8 we overlay the CO(2-1) map on the full resolution ROSAT X-ray image.  The CO(2-1) line agrees well with the soft X-ray distribution:  the X-ray shell coincides with the blue CO emission and the base of the shell with the starburst clusters.     

\citet{2012A&A...546A..65O}'s dominant hard x-ray source, X1, agrees with the brightest near-infrared clump in NICMOS images.  We identify the near-infrared peak with the brighter (northern) sub-clump in the southern starburst region of Figure 2, although the formal coordinates disagree by about 1\arcsec~; \citet{2012A&A...546A..65O} also note this inconsistency in the coordinates.  With this alignment, the secondary hard X-ray source X2 agrees with the weaker southern sub-clump.  We do not have a candidate for the X3 source that \citet{2012A&A...546A..65O}  suggest may not belong to the galaxy.   
 \begin{figure}[h]
 \begin{center}
\includegraphics*[scale=0.15]{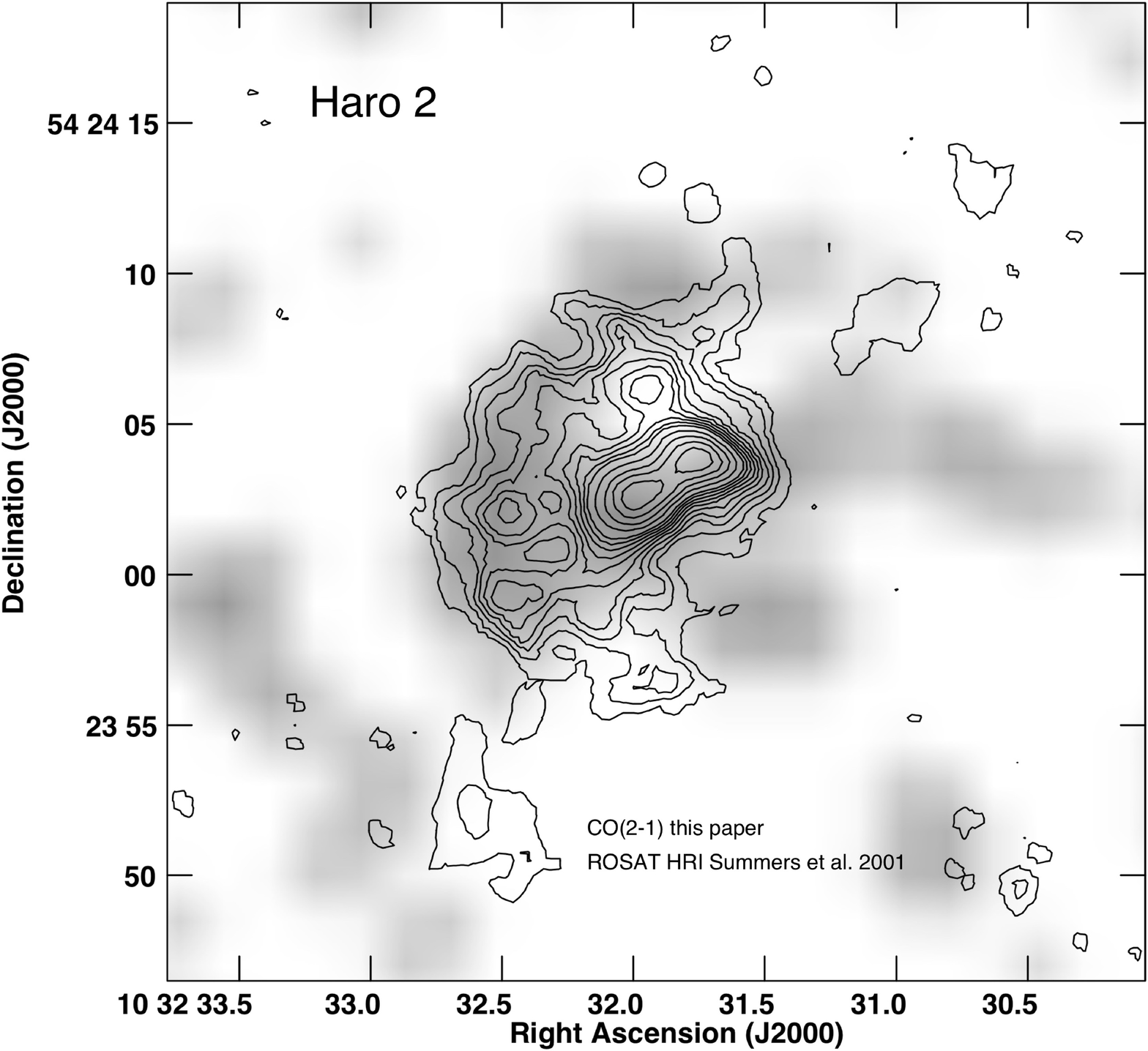}
\end{center}
\caption{The X-ray emission from the map of \citet{2001MNRAS.327..385S} in grayscale and the integrated CO(2-1) emission in contours. Contour levels are $n\times$ 600 mJy/bm km/s for $n=1-10$ and 
$n\times1200$ mJy/bm km/s  for n greater than 10.  }
\end{figure}

The CO(2-1) agrees spatially with the X-ray better than does the CO(1-0); \citet{2004AJ....127..264B} show in their Figure 8 that the X-ray extends further to the east than does the CO(1-0) emission. That the CO(2-1) is a better match than the (1-0) with the X-ray may be partly due to the improved sensitivity of the CO(2-1)  but also  because the gas is warm or hot:  a jiggle-mode maps of CO(3-2) with 14.5\arcsec~ resolution  in the JCMT Nearby Galaxies Legacy Survey \citep{2012MNRAS.424.3050W} shows (their figure C26) the CO(3-2) emission clearly extended north and east of the optical nucleus, consistent with the CO(2-1) appearance.  That the higher CO transitions are consistently stronger than the (1-0) east of the galaxy implies that the eastern gas is warm and the upper levels populated, as is seen near other strong starbursts \citep{2016ApJ...833L...6C}.
We conclude that the extended blue molecular gas, be it bubble or outflow, is associated with the hot X-ray emitting gas.   The double velocity peaks of the CO line here resembles that on an X-ray source in He~2-10, \citep{2018ApJ...867..165B} and CO lines are seen emitted from the edges of other bubbles and superbubbles (e.g. \citet{2005ApJ...618..712M}, \citet{2017ApJ...843...61S}, \citet{2009PASJ...61..237T}).  The molecular gas is probably concentrated in the thin, dense shell created in the 'snowplow' stage \citep{2001MNRAS.327..385S}.  We estimate the total energy in the outflowing molecular gas from its mass of $\approx1\times10^8 M_\odot$ (derived with $\rm X_{CO}=4\times10^{20}~\rm cm^{-2}\,(K\, km \,s^{-1})^{-1}$) and a typical 35\kms\ velocity to be $\approx10^{54}~\rm ergs$, again similar to He~2-10. 

\subsubsection{Kinematics of the X-Ray Bubble, Molecular Gas and $H\alpha$ }

  \citet{2001MNRAS.327..385S} model their X-ray bubble as resulting from the stellar winds of the nuclear starburst under the assumption that it is the same process as the ionized outflow observed in $H\alpha$.  They accordingly use the size and expansion velocity \citet{1997A&A...326..929L} found for the $H\alpha$ system to determine parameters of the the X-ray bubble.  But the current data show clearly that the X-ray and molecular bubble is {\it east} of the nucleus.   Although the spectra of \citet{1997A&A...326..929L} cannot be registered spatially because of the very poor seeing, it is reasonable to associate the ionized outflow they observe in $H\alpha$ with the $H\alpha$ filaments seen in the deep images of \citet{2000A&A...359..493M} {\it west} of the nucleus.   The velocities are another discrepancy between the ionized and molecular kinematics: the ionized outflow has expansion velocity $\sim200$\kms~,  significantly higher than the  $\sim40$\kms\ of the molecular velocities.  

We conclude that the nuclear region of \harotwo\ drives {\it two} outflows: to the west, a fast ionized outflow which has created $H\alpha$ filaments, and to the east a bubble of hot, X-ray emitting gas which has entrained warm molecular gas. 
 
How will  the results calculated by \citet{2001MNRAS.327..385S} change in this model?  The important difference is that the expansion velocity may have been overestimated. If the observed molecular velocities of $\sim40$\kms\  are used, instead of the  200\kms\ found from the $H\alpha$ results, for $v_b$ in their Equation 8, the age estimate is $1.8\times10^7$ yr, instead of $3.2\times10^6$ yr.   The longer time is more consistent than the short with other estimates of the age of the relatively mature starburst of \harotwo\ .  The longer lifetime in turn reduces the mechanical injection power derived; \citet{2001MNRAS.327..385S}'s Equation 1 gives  $1.35\times10^{39}~\rm erg~s^{-1}$ instead of $2.4\times10^{41}~\rm erg~s^{-1}$.   Such energy can be readily supplied by the stellar population of any one of the nuclear clumps. The most likely driver is in the northern source, as its velocity overlaps with velocities in the CO bubble.   

   \subsubsection{Re-evaluating the Global Kinematics of \harotwo\ }
Identifying the blue-shifted molecular gas NE of the galaxy as an outflow means that we must re-appraise the overall CO and HI kinematics described by \citet{2004AJ....127..264B}.   While the blue-shifted holds only $\sim1/2$ of the total gas mass, it is so spatially distinct that it appears clearly in first moment maps and create an apparent velocity gradient NE to SW, at an angle to the optical major axis. We believe this is what \citet{2004AJ....127..264B} see in their HI data as well.    Now that we treat the blue CO emission as a separate feature, a bubble driven by the northern star clusters, the only  velocity gradient is along the photometric major axis.  

Is the CO and X-ray bubble the base of a true galactic wind; in other words, will the matter in the bubble escape  from the galaxy?  We do not have a model for the mass distribution of \harotwo\ but can make a first estimate by treating it as an dE type; escape velocity from a dE galaxy of \harotwo\ 's luminosity is expected to be $\approx80\pm15$\kms ( \citet{ 2004ApJ...607L...9M}, Figure 2).  If this approximation is valid to within a factor of 2, it would indicate that only the very fastest gas may escape from the galaxy, while the bulk of the molecular gas will enter new orbits or fall back onto the galaxy core.

\section{Summary and Discussion}
We report observations of CO (2-1) in \harotwo\ which combine the SMA in extended and compact arrays to give beam sizes  $\approx2-7$\arcsec~.   The major findings are:
\begin{itemize}
\item Approximately $1/2$ the molecular gas is in two clumps, kinematically and spatially distinct and coincident with the bright radio sources which are identified as obscured starburst clumps.   
\item Approximately $1/2$ of the molecular gas is in a blue-shifted outflow or bubble north-east of the galactic nucleus, with expansion velocity $\sim35-40$\kms~. 
\item The blue CO outflow coincides with an X-ray superbubble which is believed to  have been created by stellar winds in the starburst, and with no other known sources.  The molecular gas has probably been entrained by the hot gas in the outflows.  
\item  The reported misalignment of the optical and kinetic axis in \harotwo\ now appears to be an artifact caused by low spatial resolution of earlier observations.
\end{itemize}

It is not clear how the stellar activity that created the molecular bubble is related to the  ionized outflow and $H\alpha$ filaments seen west of the galaxy by \citet{2000A&A...359..493M} .   Is one starburst clump driving two outflows of different types in different directions?  Are the ionized and molecular outflows driven by two different sources?   Is there an ionized outflow connected to the molecules and X-rays that has not yet been detected?    It would be very useful to observe \harotwo\ further in optical emission lines to determine the full extent of its outflow activity, and in  higher transitions of CO to determine conditions in the molecular clouds associated with the  X-ray superbubble.

\acknowledgments{SCB thanks You-Hua Chu for helpful discussions and ASIAA for hospitality during this work.}  

\vspace{5mm}
\facilities{SMA}

\software{CASA,  
          AIPS, 
          ds9
          }

\nopagebreak

\begin{table}
\caption{Observational Parameters}
\begin{tabular}{cccccc}
\tableline\tableline
\rotate
\tablewidth{0pt}
Date & Telescope & Wavelength & Beam Size & Spectral Channels & noise \\
 & & & & & $Jy/bm$ \\
\tableline
18/4/16 & SMA-Extended & 230 Ghz &$1.25\times0.97$\arcsec & $65\times 4.2$\kms & $9\times10^{-3} $ \\
04/01/16 & SMA-Compact & ''  &  $7.11\times4.55$\arcsec & '' & $1.9\times10^{-2} $ \\
n.a. &  SMA-Combined & ''  & $1.96\times1.61$\arcsec &  ''  & $1.7\times10^{-2} $ \\
02/11/90 & VLA-C array  & 4.86 Ghz & $0.59\times0.47$\arcsec & n.a. &  $8.0\times10^{-5}$ \\

\tableline
\end{tabular}
\end{table}

\clearpage
\section{APPENDIX}
\smallskip
\centerline{Kinematic Channel Maps}
\begin{figure}[ht]
\begin{center}
\includegraphics*[scale=1]{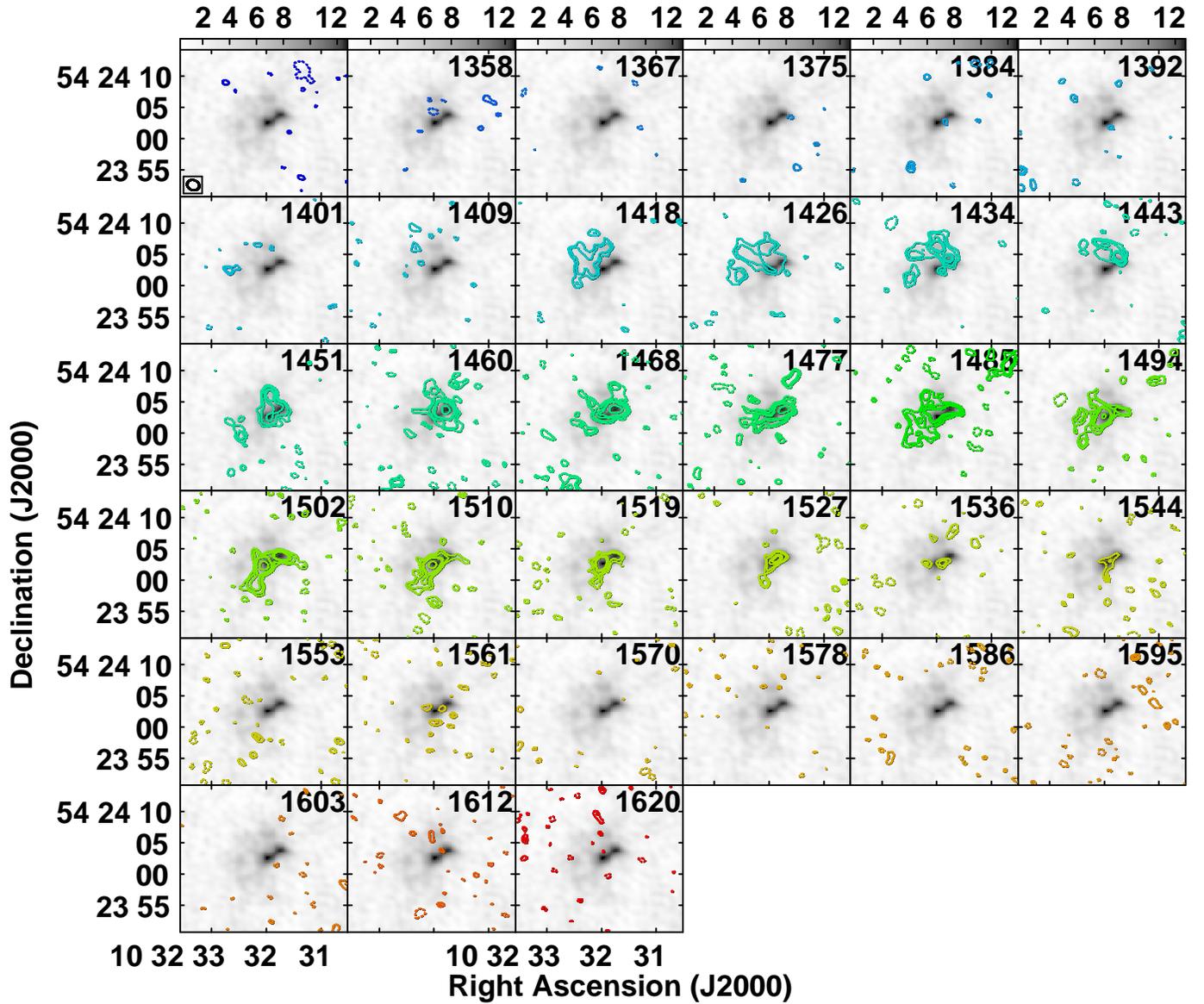} 
\caption{Velocity channel maps of \harotwo\ showing all velocity channels, overlaid on the moment 0 total emission map.  
The contours are $4\times10^{-2}\times2^{n/2}$ Jy/bm  and the greyscale range is -0.25 to 12.37 (Jy/bm)(\kms). }
\end{center}
\end{figure}

\end{document}